\documentclass{article}
\usepackage{amssymb}

\newtheorem{theorem}{Theorem}

\newtheorem{definition}[theorem]{Definition}

\newtheorem{remark}[theorem]{Remark}

\input{tcilatex}
\begin{document}

\title{\textbf{Newton's 2}$nd$\textbf{\ Law, \ Radiation \ Reaction \& Type
II \ \ Einstein-Maxwell Fields\ }\ \ }
\date{ \ .}
\author{ {\tiny .\ .\ }\ \ \ \ \ \ \ \ \ \ \ \ \ Ezra T Newman \\
\ {\tiny .\ }\ \ \ \ \ \ \ \ \ Dept of Physics and Astronomy, \\
\ {\tiny .\ \ }\ \ \ Univ of Pittsburgh, Pittsburgh PA 15260}
\date{ . \ \ \ \ \ \ \ \ \ \ \ \ \ \ \ 9/18/11}
\maketitle

\begin{abstract}
Considering perturbations off the Reissner-Nordstrom metric while keeping
the perturbations in the class of type II Einstein-Maxwell \ \ \ metrics, we
do a spherical harmonic expansion of all the variables up to the quadrupole
term. \ This leads to a rather surprising results. \ Referring to the source
of the metric as a type II particle (analogous to referring to a
Schwarzschild-Reissner-Nordstrom or Kerr-Newman particle), we see
immediately that the Bondi momentum of the particle take the classical form
of mass times velocity plus an electromagnetic radiation reaction term while
the Bondi mass loss equation become the classical gravitational and
electromagnetic (electric and magnetic) dipole and quadrupole radiation. The
Bondi momentum loss equation turns into Newtons second law of motion
containing the Abraham, Lorentz, Dirac radiation reaction force plus a
momentum recoil (rocket) force while the reality condition on the Bondi mass
aspect yields the conservation of angular momentum.
\end{abstract}

\section{ Introduction}

For a variety of reasons the algebraically special Einstein or
Einstein-Maxwell metrics and equations, \cite%
{Pirani,GS,RT,Kinnersley,UCF,Lind,Talbot,NT}\textbf{\ }have played a major
role in general relativity for many years: they have contained some of the
most studied and useful solutions of the Einstein-Maxwell equations
(Schwarzschild, Reissner-Nordstrom, Kerr-Newman\textbf{\cite{KN}, }the
Robinson-Trautman metrics\cite{RT}, plane gravitational waves), they contain
the beautiful Goldberg-Sachs theorem\cite{GS} with its emphasis on the
importance of shear-free null geodesic congruences and are the source of
much attractive mathematical research, e.g., existence theorems, CR manifold
theory.

In this work we will be concerned only with the type II field equations and
their solutions - and some very surprising physical results that have been
hiding in them - results that have been overlooked for many years. Though
the field equations are for the metric variables we will adopt a slightly
unconventional point of view. In analogy with referring to the source of a
Schwarzschild, Reissner-Nordstrom or Kerr-Newman metric as a Schwarzschild,
Reissner-Nordstrom or Kerr-Newman particle, we will refer to the source of
the type II metrics as type II particles. \textit{The field equations are
then interpreted as the equations governing the dynamics (motion and
multipole behavior) of the type II particles. }The metric itself is given
here for completeness, but it and its immediate geometric properties are not
used.

If one starts with the full Einstein-Maxwell equations and imposes the type
II conditions many of the resulting equations (the radial equations) can be
easily integrated, leaving four (complex) \textit{reduced }type II
Einstein-Maxwell equations - non-linear and rather intractable looking - for
four complex field variables, which are functions of time and two angles,
plus a reality condition. Two of the four dependent variables are
spin-coefficient versions of the Maxwell field, one of them is a
spin-coefficient Weyl tensor component and the last is a (geometric)
direction field that specifies the direction (at null infinity) of the
shear-free null geodesic congruence defined by the type II condition. The
structure of the four field equations is as follows: the first two equations
and the third determine respectively, the two Maxwell fields and the Weyl
tensor component (that contains the Bondi energy-momentum four-vector) in
terms of the direction field. The last equation, basically a conservation
law, is the dynamics for the direction field.

Working within the algebraically special type II Einstein-Maxwell field
equations, we study perturbations from the Reissner-Nordstrom metric.
Specifically we consider linear perturbations (for the first three
equations) with spherical harmonic expansions up to and including the $l=2$
terms. \ The fourth of the field equations, (the $l=0,1$ terms form the
Bondi energy-momentum conservation law) is intrinsically quadratic and is
essentially empty if linearized. With no inconsistency, the results obtained
from the three linearized equations are inserted into the fourth. (This is
analogous to first solving the \textit{linear} Maxwell equations and then
using these solutions in the \textit{quadratic} stress-tensor conservation
laws.) Our main results are in the $l=0\ $and$\ $1$\ $harmonics of the last
equation. \ The $l=0\ $is the energy conservation law containing both
gravitational quadrupole radiation and electric and magnetic dipole and
quadrupole radiation in complete agreement with known classical results. The 
$l=1$\ terms yield Newtons 2$^{nd}\ \ $law of motion ($F=Ma$), with the
force coming from both radiation reaction terms and momentum recoil terms.
With the momentum recoil terms it is a generalization of the Abraham,
Lorentz, Dirac equation. \ The $l=2\ $terms determine the evolution of the
quadrupole moments. \textit{It must be emphasized that these results do not
involve any model building and do not contain any mass renormalization
procedure. They are simply results that have been sitting there in the type
II equations waiting to be seen.}

In section II, we first define and then discuss the type II metrics and
Maxwell fields with their relevant variables, their four differential
equations and reality condition. In addition we review the theory of
shear-free null geodesic congruences (NGCs) and their associated complex
world-lines. It is the theory of NGCs that lead to the basic variables, the
direction field as well as the particle position vector. Section III is
devoted to the approximations: initially the linearization of the first
three of the field equations, then their expansion in the three spherical
harmonics, $l=0,1,2,$ and finally their integration. The physical results
and their interpretation are discussed in sections III and IV.

\section{Background}

\subsection{Notation}

Much of our analysis takes place on the future null infinity, $\mathfrak{I}%
^{+},\ $which is coordinatized by Bondi coordinates: the null generators, by
the complex stereographic coordinates, ($\zeta ,\overline{\zeta }$) and the
slices of $\mathfrak{I}^{+}$ by $u.\ $Though we are interested only in real
results we must consider, along the way, the complexification of $\mathfrak{I%
}^{+}.\ $In that case we allow $u$ to take complex values and consider $%
\overline{\zeta }\ $to be close to but independent of$\ $the complex
conjugate\ of $\zeta $.

Later we will switch from the Bondi $u$\ to $\tau =\sqrt{2}c^{-1}u\ $with
the notation change: $\partial _{u}W=W^{{\large \cdot }}=\sqrt{2}%
c^{-1}\partial _{\tau }W\equiv \sqrt{2}c^{-1}W^{\prime }$

We use the Lorentzian tetrad for the angular behavior of the field variables:

\begin{eqnarray}
\hat{l}^{a} &=&\frac{\sqrt{2}}{2(1+\zeta \overline{\zeta })}\left( 1+\zeta 
\overline{\zeta },\zeta +\overline{\zeta },-i(\zeta -\overline{\zeta }%
),-1+\zeta \overline{\zeta }\right) =(\frac{\sqrt{2}}{2},\frac{1}{2}%
Y_{1i}^{0}),  \label{nulltetrad1} \\
\hat{m}^{a} &=&\frac{\sqrt{2}}{2(1+\zeta \overline{\zeta })}\left( 0,1-%
\overline{\zeta }^{2},-i(1+\overline{\zeta }^{2}),2\overline{\zeta }\right)
=(0,-Y_{1i}^{1}),  \label{2} \\
\overline{\hat{m}}^{a} &=&\frac{\sqrt{2}}{2(1+\zeta \overline{\zeta })}%
\left( 0,1-\zeta ^{2},i(1+\zeta ^{2}),2\zeta \right) =(0,-Y_{1i}^{-1}),
\label{3} \\
\hat{n}^{a} &=&\frac{1}{2(1+\zeta \overline{\zeta })}\left( 1+\zeta 
\overline{\zeta },-(\zeta +\overline{\zeta }),i(\zeta -\overline{\zeta }%
),1-\zeta \overline{\zeta }\right) =(\frac{\sqrt{2}}{2},-\frac{1}{2}%
Y_{1i}^{0}).  \label{4}
\end{eqnarray}

\subsection{Algebraic Types}

In every Lorentzian space-time there are four null vectors (principal null
vectors, [pnv], $L^{a}$) per space-time point that are determined by
solutions to the algebraic equation\cite{Pirani}%
\begin{equation}
L^{b}L_{[e}C_{a]bc[d}L_{f]}L^{d}=0,  \label{Pirani}
\end{equation}%
with $C_{abcd}\ $being the Weyl tensor. In the case when two or more of
these vectors are degenerate (or coincide) the associated metric is referred
to as algebraically special. We consider only the double degeneracy case
referred to as type II. (Other types are called types III, IV and D.) In the
NP formalism this condition translates to the requirement that a null tetrad
system exists such that the spin-coefficient Weyl tensor components satisfy $%
\psi _{0}\ $= $\psi _{1}=0$ and the Maxwell component $\phi _{0}=0.\ $It
then follows from the beautiful Goldberg-Sachs\cite{GS} theorem that the
degenerate pnv's are the tangent vectors to a null geodesic congruence that
is \textit{shear-free}. These conditions greatly simplify the integration of
the Einstein-Maxwell equations. \ 

\subsection{Type II Metric and Field Equations}

Though it plays no role in what follows other than to define our dependent
variables, for completeness we display the type II metric\cite{NT,Lind}:%
\begin{equation}
g=2(l^{^{\ast }}n^{\ast }+n^{\ast }l^{\ast }-m^{\ast }\overline{m}^{\ast }-%
\overline{m}^{\ast }m^{\ast }),  \label{metric*}
\end{equation}%
with

\begin{eqnarray}
l^{\ast } &=&du-P^{-1}(Ld\zeta +\overline{L}d\overline{\zeta }),
\label{tetrad*} \\
m^{\ast } &=&-P^{-1}[r-i\Sigma ]d\overline{\zeta }=P^{-1}\overline{\rho }%
^{-1}d\overline{\zeta },  \nonumber \\
\overline{m}^{\ast } &=&-P^{-1}[r+i\Sigma ]d\zeta =P^{-1}\rho ^{-1}d\zeta , 
\nonumber \\
n^{\ast } &=&dr+P^{-1}[(K+L^{{\large \cdot }}\overline{\rho }^{-1})d\zeta +(%
\overline{K}+\overline{L}^{{\large \cdot }}\rho ^{-1})d\overline{\zeta }] 
\nonumber \\
&&+[\frac{1}{2}(\rho \psi _{2}^{\ast 0\,}+\overline{\rho }\overline{\psi }%
_{2}^{\ast 0\,})+(1+\frac{1}{2}[\eth \overline{L}^{{\large \cdot }}+\text{ }%
\overline{\eth }L^{{\large \cdot }}+L\overline{L}^{{\large \cdot \cdot }}+%
\overline{L}L^{{\large \cdot \cdot }}])]l^{\ast },  \nonumber
\end{eqnarray}

where 
\begin{eqnarray}
\rho  &=&-\frac{1}{r+i\Sigma },  \label{rho} \\
i\Sigma (u,\zeta ,\overline{\zeta }) &\equiv &\frac{1}{2}(\eth \overline{L}+L%
\overline{L}^{{\large \cdot }}-\overline{\eth }L-\overline{L}L^{{\large %
\cdot }}),  \label{sigma}
\end{eqnarray}%
\begin{eqnarray}
K &=&L+L\eth \overline{L}^{{\large \cdot }}+\frac{1}{2}(\eth ^{2}\overline{L}%
+L^{2}\overline{L}^{{\large \cdot \cdot }}+L^{{\large \cdot }}[\eth 
\overline{L}-\overline{\eth }L+L\overline{L}^{{\large \cdot }}-\overline{L}%
L^{{\large \cdot }}]),  \label{K} \\
P &=&1+\zeta \overline{\zeta },
\end{eqnarray}%
with $\rho $ being the complex divergence of the shear-free NGC and $\Sigma $
its twist. The coordinates are ($r,u,\zeta ,\overline{\zeta }$), with $r\ $%
being the radial coordinate (affine, along the type II NGC) and ($u,\zeta ,%
\overline{\zeta }$) the Bondi coordinates on future null infinity,$\ 
\mathfrak{I}^{+}.$

The spin-coefficient components of the associated Maxwell field in the $%
l^{\ast },n^{\ast },m^{\ast },\overline{m}^{\ast }\ $tetrad are 
\begin{eqnarray}
F_{ab}l^{\ast a}m^{\ast b} &=&\phi _{0}^{\ast }=0,  \label{max0} \\
\frac{1}{2}F_{ab}(l^{\ast a}n^{\ast b}+\overline{m}^{\ast a}m^{\ast b})
&=&\phi _{1}^{\ast }=\frac{\phi _{1}^{\ast 0}}{r^{2}}+O(r^{-3}),
\label{max1} \\
F_{ab}\overline{m}^{\ast a}n^{\ast b} &=&\phi _{2}^{\ast }=\frac{\phi
_{2}^{\ast 0}}{r^{2}}+O(r^{-2}).  \label{max2}
\end{eqnarray}%
\qquad 

Of relevance to us is that our `particle' variables are seen to be the
complex $L(u,\zeta ,\overline{\zeta }),\ \psi _{2}^{\ast 0\,}(u,\zeta ,%
\overline{\zeta })\ $coming from the metric and the two Maxwell variables, $%
\phi _{1}^{\ast 0}(u,\zeta ,\overline{\zeta }),\ \phi _{2}^{\ast 0}(u,\zeta ,%
\overline{\zeta }).$

It has been shown \cite{Lind,Talbot}, that they satisfy the following set of
four field equations: 
\begin{eqnarray}
\eth \phi _{1}^{\ast 0}+2L^{{\large \cdot }}\phi _{1}^{\ast 0}+L(\phi
_{1}^{\ast 0})^{{\large \cdot }} &=&0,  \label{m1} \\
(\phi _{1}^{\ast 0})^{{\large \cdot }}+\eth \phi _{2}^{\ast 0}+(L\phi
_{2}^{\ast 0})^{{\large \cdot }} &=&0,  \label{m2}
\end{eqnarray}

\begin{eqnarray}
\eth \psi _{2}^{\ast 0} &=&-L(\psi _{2}^{\ast 0\,})^{{\large \cdot }}-3L^{%
{\large \cdot }}\psi _{2}^{\ast 0}+2k\phi _{1}^{\ast 0}\overline{\phi }%
_{2}^{\ast 0},  \label{gr1} \\
\Psi ^{{\large \cdot }} &=&\sigma ^{{\large \cdot }}(\overline{\sigma })^{%
{\large \cdot }}+k\phi _{2}^{\ast 0}\overline{\phi }_{2}^{\ast 0},
\label{gr2}
\end{eqnarray}%
with the \textit{mass aspect} given by

\begin{definition}
\begin{equation}
\Psi \equiv \psi _{2}^{\ast 0}+2L\eth (\overline{\sigma })^{{\large \cdot }%
}+L^{2}(\overline{\sigma })^{{\large \cdot \cdot }}+\eth ^{2}\overline{%
\sigma }+\sigma (\overline{\sigma })^{{\large \cdot }},  \label{mass aspect}
\end{equation}%
(with the very important reality condition),%
\begin{equation}
\Psi =\overline{\Psi }.  \label{reality}
\end{equation}%
and%
\begin{eqnarray}
\sigma &\equiv &\eth L+LL^{{\large \cdot }}  \label{S} \\
k &=&2Gc^{-4},  \label{k'} \\
L^{{\large \cdot }} &\equiv &\frac{\partial L}{\partial u},\ etc.
\label{dott}
\end{eqnarray}
\end{definition}

Note that we can switch back and forth between the mass aspect $\Psi \ $and
the Weyl component, $\psi _{2}^{\ast 0}$, via Eq.(\ref{mass aspect}) as
needed.

In section III we will take these equations, partially linearize them,
expand them in spherical harmonics and then analyze and display their
physical content.

\subsection{Shear-Free Null Geodesic Congruences \& Complex World-lines}

The evolution of \textit{generic asymptotically flat Einstein-Maxwell
space-times} (described in Bondi coordinates and tetrad, $l,n,m,\overline{m}$%
) are driven by the freely given asymptotic shear, $\sigma $($u,\zeta ,%
\overline{\zeta }$), of the associated Bondi null geodesic congruence (NGC),
(the generators of the Bondi null surfaces) and Maxwell radiation field. In
addition to the Bondi NGC, other NGCs, defined by the rotation of the
tangent vector of the Bondi null geodesic congruence, can be introduced.
They in turn come with an associated transformed shear. More specifically,
at each point ($u,\zeta ,\overline{\zeta }$) of $\mathfrak{I}^{+},$we
consider its past light cone and its sphere of null direction with the
angles given by the pair of complex stereographic coordinates, ($L,\overline{%
L}$). (The north and south poles of the sphere are fixed by the null
generators of $\mathfrak{I}^{+}\ $and the Bondi vector $l.$) Any field of
null directions on $\mathfrak{I}^{+},\ ($which automatically determines an
interior NGC), can be given by some complex regular spin-weight-one function 
$L=L(u,\zeta ,\overline{\zeta }).$ From a given Bondi tetrad, ($%
l^{a},n^{a},m^{a},\overline{m}^{a}$), a second tetrad, ($l^{\,\ast
},n^{\,\ast },m^{\,\ast },\overline{m}^{\,\ast }),\ $in the neighborhood of $%
\mathfrak{I}^{+}$, can be constructed by the null rotation, with arbitrary $%
L(u,\zeta ,\overline{\zeta }):$

\begin{eqnarray}
l^{a} &\rightarrow &l^{\,\ast a}=l^{a}-\frac{\bar{L}}{r}m^{a}-\frac{L}{r}%
\bar{m}^{a}+0(r^{-2}),  \label{null rot} \\
m^{a} &\rightarrow &m^{\ast a}=m^{a}-\frac{L}{r}n^{a}+0(r^{-2}), \\
n^{a} &\rightarrow &n^{\ast a}=n^{a}+0(r^{-2}).
\end{eqnarray}

The asymptotic shear of the field of new tangent null vectors $l^{\,\ast a}\ 
$is given by \cite{Aronson} 
\begin{equation}
\sigma ^{0\ast }=\sigma ^{0}-\eth L-LL^{\cdot }  \label{sigma*}
\end{equation}%
so that \textit{shear-free} NGC's, (from $\sigma ^{0\ast }=0$), are
determined by functions $L(u,\zeta ,\overline{\zeta })\ $that satisfy the
differential equation

\begin{equation}
\eth L+LL^{\cdot }=\sigma ^{0}.  \label{shear-free}
\end{equation}

In other words a function $L(u,\zeta ,\overline{\zeta }),\ $that satisfies
Eq.(\ref{shear-free}), \textit{determines an asymptotically shear-free} NGC.

It has been shown\cite{Footprints,LR} that \textit{regular\ solutions }to%
\textit{\ }Eq.(\ref{shear-free}) are generated by arbitrary complex analytic
curves in an auxiliary space referred to as $\mathfrak{H}$-space\cite%
{H-Space}. One begins by finding solutions to the `good-cut' equation

\begin{equation}
\eth ^{2}Z=\sigma ^{0}(Z,\zeta ,\overline{\zeta })  \label{GCEq}
\end{equation}%
which are known to depend on four complex parameters, the coordinates of $%
\mathfrak{H}$-space, $z^{a}$. The solution (in general complex), written as 
\begin{equation}
u=Z(z^{a},\zeta ,\overline{\zeta }),  \label{good cuts}
\end{equation}%
is interpreted as describing a \textit{four-parameter family of slices}
(cuts) of complexified $\mathfrak{I}^{+},\ $the so-called `good-cuts'. By
choosing an \textit{arbitrary} complex analytic curve in $\mathfrak{H}$%
-space, parametrized by complex $\tau $, $z^{a}=\xi ^{a}(\tau ),\ $we have a
one-complex parameter family of slices,%
\begin{equation}
u=Z(\xi ^{a}(\tau ),\zeta ,\overline{\zeta })=G(\tau ,\zeta ,\overline{\zeta 
}).  \label{good-cuts}
\end{equation}%
The \textit{regular} solutions to Eq.(\ref{shear-free}), $L(u,\zeta ,%
\overline{\zeta }),$ are the stereographic angle field determined from the
null normals to these slices. \ The solutions have the parametric form%
\begin{eqnarray}
L &=&\eth _{(\tau )}G(\tau ,\zeta ,\overline{\zeta }),
\label{parametric sol.I} \\
u &=&G(\tau ,\zeta ,\overline{\zeta }),  \label{parametric sol.2}
\end{eqnarray}%
the subscript ($\tau $) indicating that the derivative is taken at constant $%
\tau .$

Shear-free and asymptotically shear-free null geodesic congruences$\mathfrak{%
\ }$that are \textit{regular, (i.e., no generators on }$\mathfrak{I}^{+}$%
\textit{) are }induced by arbitrary complex curves in $\mathfrak{H}$-space%
\cite{Footprints,LR}.

These results are applied to our considerations of asymptotically flat type
II metrics by the following observation: Among all the \textit{%
asymptotically shear-free} NGCs that exist in a type II space-time, there is
a \textit{unique one}, determined by the Goldberg-Sachs\cite{GS} theorem for
the degenerate principal null vector field, that is not only \textit{%
asymptotically shear-free }but also\textit{\ totally} shear free (aside from
caustic regions). \ It is this particular complex world-line, with its
associated cut-function, Eq.(\ref{good-cuts}) that govern - or act as the
backbone - to the solutions to the field equations, (\ref{m1}) -(\ref{gr2}).

Anticipating the approximations of the next section, the spherical harmonic
expansion of the function, $u=G(\tau ,\zeta ,\overline{\zeta })\ \ $and the
derived $\sigma ^{0}(u,\zeta ,\overline{\zeta }),\ L(u,\zeta ,\overline{%
\zeta }),$ (parametrically described) up to the $l=2\ $harmonic$,\ $are:%
\begin{eqnarray}
u &=&G(\tau ,\zeta ,\overline{\zeta })=\xi ^{a}(\tau )\hat{l}_{a}(\zeta ,%
\overline{\zeta })+\xi ^{ij}(\tau )Y_{2ij}^{0}  \label{G} \\
&=&\tau -\frac{1}{2}\xi ^{i}Y_{1i}^{0}(\zeta ,\overline{\zeta })+\xi
^{ij}(\tau )Y_{2ij}^{0},  \nonumber \\
L(u,\zeta ,\overline{\zeta }) &=&\eth _{(\tau )}G(\tau ,\zeta ,\overline{%
\zeta })=\xi ^{i}(\tau )Y_{1i}^{1}-6\xi ^{ij}(\tau )Y_{2ij}^{1},  \label{L}
\\
\sigma ^{0}(u,\zeta ,\overline{\zeta }) &=&\eth _{(\tau )}^{2}G(\tau ,\zeta ,%
\overline{\zeta })=24\xi ^{ij}(\tau )Y_{2ij}^{2}.  \label{SIGMA}
\end{eqnarray}

The last of the field equations become the evolution equations for the $\xi
^{i}(\tau )\ $and $\xi ^{ij}(\tau ).$

\section{The Field Equations}

\subsection{Linearization and Harmonic Expansion}

We begin with the field equations\cite{Lind}, (\ref{m1})-(\ref{reality}),

\begin{eqnarray}
\eth \phi _{1}^{\ast 0} &=&-2L^{{\large \cdot }}\phi _{1}^{\ast 0}-L(\phi
_{1}^{\ast 0})^{{\large \cdot }},  \label{a} \\
(\phi _{1}^{\ast 0})^{{\large \cdot }} &=&-\eth \phi _{2}^{\ast 0}-(L\phi
_{2}^{\ast 0})^{{\large \cdot }},  \label{b} \\
\eth \psi _{2}^{\ast 0} &=&-L(\psi _{2}^{\ast 0\,})^{{\large \cdot }}-3L^{%
{\large \cdot }}\psi _{2}^{\ast 0}+4Gc^{-4}\phi _{1}^{\ast 0}\overline{\phi }%
_{2}^{\ast 0},  \label{c} \\
\Psi &=&\overline{\Psi }\equiv \psi _{2}^{\ast 0}+2L\eth (\overline{\sigma }%
)^{{\large \cdot }}+L^{2}(\overline{\sigma })^{{\large \cdot \cdot }}+\eth
^{2}\overline{\sigma }+\sigma (\overline{\sigma })^{{\large \cdot }},
\label{d} \\
\Psi ^{{\large \cdot }} &=&\sigma ^{{\large \cdot }}(\overline{\sigma })^{%
{\large \cdot }}+2Gc^{-4}\phi _{2}^{\ast 0}\overline{\phi }_{2}^{\ast 0},
\label{e} \\
\sigma &\equiv &\eth L+LL^{{\large \cdot }},  \label{f}
\end{eqnarray}%
and first note that the Reissner-Nordstrom solution is given by%
\begin{eqnarray}
\phi _{1}^{\ast 0} &=&q, \\
\phi _{0}^{\ast 0} &=&\phi _{2}^{\ast 0}=L=0, \\
\psi _{2}^{\ast 0} &=&\Psi =\Psi ^{0}=-M_{B}\frac{2\sqrt{2}G}{c^{2}}.
\end{eqnarray}%
with the zero order terms, $q\ $and $M_{B},$being the charge and Bondi mass$%
. $

The linearization of the first four equations, (\ref{a})-(\ref{d}), leads to%
\begin{eqnarray}
\eth \phi _{1}^{\ast 0} &=&-2qL^{{\large \cdot }},  \label{a'} \\
(\phi _{1}^{\ast 0})^{{\large \cdot }} &=&-\eth \phi _{2}^{\ast 0},
\label{b'} \\
\eth \psi _{2}^{\ast 0} &=&6\sqrt{2}Gc^{-2}M_{B}L^{{\large \cdot }}+4Gc^{-4}q%
\overline{\phi }_{2}^{\ast 0},  \label{c'} \\
\Psi &=&\overline{\Psi }\equiv \psi _{2}^{\ast 0}+\eth ^{2}\overline{\eth }%
\overline{L}^{{\large \cdot }}.  \label{d'}
\end{eqnarray}

By eliminating $\psi _{2}^{\ast 0}\ $in terms of $\Psi ,\ $via (\ref{d'}),
inserting $c\ $explicitly and rescaling $u\ \ $(to obtain retarded time) by $%
u=\frac{\sqrt{2}}{2}c\tau ,\ $(i.e.,$\ $define $K^{{\large \cdot }}=\sqrt{2}%
c^{-1}\partial _{\tau }K\equiv \sqrt{2}c^{-1}K^{\prime }$)$,\ $we obtain our
three linear equations:

\begin{eqnarray}
\eth \phi _{1}^{\ast 0} &=&2q\sqrt{2}c^{-1}L^{\prime },  \label{a''} \\
\phi _{1}^{\ast 0\prime } &=&-\frac{\sqrt{2}}{2}c\eth \phi _{2}^{\ast 0},
\label{b''} \\
\eth \Psi &=&12Gc^{-3}M_{B}L^{\prime }+\sqrt{2}c^{-1}\eth ^{3}\overline{\eth 
}\overline{L}^{\prime }-4Gc^{-4}q\overline{\phi }_{2}^{\ast 0}.  \label{c''}
\end{eqnarray}

Using the spherical harmonic expansions,%
\begin{eqnarray}
L(u,\zeta ,\overline{\zeta }) &=&\eth _{(\tau )}G(\tau ,\zeta ,\overline{%
\zeta })=\xi ^{i}(\tau )Y_{1i}^{1}-6\xi ^{ij}(\tau )Y_{2ij}^{1},  \label{A}
\\
\phi _{1}^{\ast 0} &=&q+\phi _{1}^{\ast 0i}Y_{1i}^{0}+\phi _{1}^{\ast
0ij}Y_{2ij}^{0}  \label{B} \\
\phi _{2}^{\ast 0} &=&-2c^{-2}D_{e\&m}^{i\prime \prime }Y_{1i}^{-1}-\frac{%
\sqrt{2}}{12}c^{-3}Q_{e\&m}^{ij\prime \prime \prime }Y_{2ij}^{-1}  \label{C}
\end{eqnarray}%
(where $D_{e\&m}^{i}\equiv D_{elect}^{i}+iD_{mag}^{i}\ $and $%
Q_{e\&m}^{ij}\equiv Q_{elect}^{ij}+iQ_{mag}^{ij}\ $are respectively the
complex dipole and complex quadrupole moments of the source, defined from
the radiation field, i.e., the $r^{-1}\ \ $part of $F^{ab}$), we find, from (%
\ref{a''}) and (\ref{b''}), with Eq.(\ref{C}), the full asymptotic Maxwell
field,

\begin{equation}
\ \phi _{1}^{\ast 0}{}=q+\sqrt{2}c^{-1}D_{e\&m}^{i\prime }(u_{r})Y_{1i}^{0}+%
\frac{1}{12}c^{-2}Q_{e\&m}^{ij\prime \prime }Y_{2ij}^{0}  \label{phi_1}
\end{equation}%
with

\begin{eqnarray}
\text{ \ }D_{e\&m}^{i} &=&q\xi ^{i},  \label{D_C} \\
\ \ Q_{e\&m}^{ij\prime } &=&-24\sqrt{2}cq\xi ^{ij}.  \label{Q_C}
\end{eqnarray}%
Note that they are given in terms of the harmonic components of $L(u,\zeta ,%
\overline{\zeta }).$

By substituting these results into the last of the linear equations, (\ref%
{c''}), we obtain for the $l=1,2\ $harmonic coefficients:%
\begin{eqnarray}
\Psi ^{i} &=&\overline{\Psi }^{i}=\frac{3\sqrt{2}}{2}c^{-1}\Psi ^{0}\xi
^{i\prime }+4Gc^{-5}q\overline{D}_{e\&m}^{i\prime \prime }\ ,  \label{PSI^i}
\\
\Psi ^{ij} &=&\overline{\Psi }^{ij}=-24\overline{\xi }^{ij}-3\sqrt{2}\Psi
^{0}c^{-1}\xi ^{ij\prime }+Gqc^{-7}\frac{\sqrt{2}}{18}\overline{Q}%
_{e\&m}^{ij\prime \prime \prime }.  \label{PSI^ij}
\end{eqnarray}

From the definition of the Bondi energy-momentum vector, ($M_{B}c^{2},P^{i}$%
);%
\begin{eqnarray}
\Psi ^{0} &=&-M_{B}\frac{2\sqrt{2}G}{c^{2}},  \label{bondi M} \\
\Psi ^{i} &=&-\frac{6G}{c^{3}}P^{i},  \label{bondi momentum}
\end{eqnarray}%
the reality conditions and the decompositions, 
\begin{eqnarray}
\xi ^{a} &=&\xi _{R}^{a}+i\xi _{I}^{a},  \label{r&I,I} \\
\xi ^{ij} &=&\xi _{R}^{ij}+i\xi _{I}^{ij}  \label{r&I.2}
\end{eqnarray}%
we find from Eq.(\ref{PSI^i}),%
\begin{eqnarray}
P^{i} &=&M_{B}\xi _{R}^{i\prime }-\frac{2}{3}c^{-2}q^{2}\xi _{R}^{i\prime
\prime }\ ,  \label{P} \\
0 &=&M_{B}\xi _{I}^{i\prime }+\frac{2}{3}c^{-2}q^{2}\xi _{I}^{i\prime \prime
}.  \label{AngularMomCons}
\end{eqnarray}

\begin{remark}
These relations have immediate physical interpretation: \ Eq.(\ref{P})
defines the Bondi momentum with the \textit{standard} $Mv$ term plus the
radiation reaction term that later becomes the well-known radiation reaction
force. \ Eq.(\ref{AngularMomCons}), rewritten as%
\begin{eqnarray}
J^{i} &=&M_{B}c\xi _{I}^{i}+\frac{2}{3}c^{-1}q^{2}\xi _{I}^{i\prime },
\label{J} \\
J^{i\prime } &=&0,  \label{J'=0}
\end{eqnarray}%
generalizes the spin-angular momentum of the Kerr-Newman metric, $%
J^{i}=M_{B}c\xi _{I}^{i}$,\ and its conservation. \ In addition the electric
and magnetic dipole moments are given, via Eq.(\ref{D_C}), by%
\begin{eqnarray}
\text{\ }D_{e}^{i} &=&q\xi _{R}^{i},  \label{D_e} \\
\text{\ }D_{m}^{i} &=&q\xi _{I}^{i}.  \label{D_m}
\end{eqnarray}
\end{remark}

The real and imaginary parts of $\Psi ^{ij}\ $(with the reality condition),
from (\ref{PSI^ij}), yield%
\begin{eqnarray}
\text{\ }\Psi ^{ij} &=&-24\xi _{R}^{ij}-3\sqrt{2}\Psi ^{0}c^{-1}\xi
_{R}^{ij\prime }-Gc^{-6}q^{2}\frac{8}{3}\xi _{R}^{ij\prime \prime },
\label{PSI^ijR} \\
0 &=&24\xi _{I}^{ij}-3\sqrt{2}\Psi ^{0}c^{-1}\xi _{I}^{ij\prime }+\frac{8}{3}%
Gc^{-6}q^{2}\xi _{I}^{ij\prime \prime }.  \label{PSI^ijI}
\end{eqnarray}

\textbf{NOTE: }Returning to the issue of the structure of our equations, it
can be seen that our linear equations have determined three of the four
dependent variables, i.e., $\phi _{1}^{\ast 0},\phi _{2}^{\ast 0}\ $and $%
\Psi \ ($or$\ \psi _{2}^{0}),$ in terms of $L\ ($or $\xi ^{i}\ $and $\xi
^{ij})$. The evolution of the \textit{imaginary parts} of $\xi ^{i}\ $and $%
\xi ^{ij}\ ($arising from the reality condition$)\ $are determined by Eqs.(%
\ref{J'=0}) and (\ref{PSI^ij})$.\ $The evolution of the real parts will be
determined by the non-linear Eq.(\ref{e}).

\subsection{The Quadratic Field Equation and Conservation Laws}

Since the linearization of Eq.(\ref{e}) yields the trivial, $\Psi ^{{\large %
\cdot }}=0,\ $we can, with no inconsistency, consider its full quadratic
version 
\begin{equation}
\sqrt{2}\Psi ^{{\large \prime }}=2\eth L^{\prime }\ \overline{\eth }%
\overline{L}^{\prime }+Gc^{-4}\phi _{2}^{\ast 0}\overline{\phi }_{2}^{\ast
0},  \label{quadratic}
\end{equation}%
where the linear results, with the Bondi energy-momentum relations, are
inserted. The procedure, essentially straightforward, (utilizing the
Clebsch-Gordon expansion of spherical harmonic products), yields for the $%
l=0,1,2\ $harmonics:

I. $l=0,$

\begin{eqnarray}
E^{\prime } &\equiv &M_{B}^{\prime }c^{2}=-\frac{1}{5}c^{-7}\boldsymbol{G}Q_{%
\mathrm{Grav}}^{ij\,\prime \prime \prime }\overline{Q}_{\mathrm{Grav}%
}^{ij\,\prime \prime \prime }  \label{energy cons} \\
&&-\frac{2}{3}c^{-3}D_{e\&m}^{i\prime \prime }\overline{D}_{e\&m}^{i\prime
\prime }-\frac{1}{180}c^{-5}Q_{e\&m}^{ij\prime \prime \prime }\overline{Q}%
_{e\&m}^{ij\prime \prime \prime }.  \nonumber
\end{eqnarray}

The first term is the classical Bondi gravitational quadrupole energy loss
while the second and third terms are the classical electromagnetic dipole
and quadrupole radiation losses\cite{LL}. Note that they contain both the 
\textit{electric and magnetic} dipole and quadrupole radiation and
furthermore, each of these expressions is given in terms of the angle fields 
$L\ $via%
\begin{eqnarray}
Q_{Grav}^{ij\ \prime \prime } &=&12\sqrt{2}G^{-1}c^{4}\xi ^{ij} \\
\text{ \ }D_{e\&m}^{i} &=&q\xi ^{i}, \\
\ \ Q_{e\&m}^{ij\prime } &=&-24\sqrt{2}cq\xi ^{ij}.
\end{eqnarray}

II. $l=1,$%
\begin{eqnarray}
P^{i\prime } &=&-i\epsilon _{ijk}(\frac{2}{15c^{4}}\boldsymbol{G}Q_{\mathrm{%
Grav}}^{kl\,\prime \prime \prime }\overline{Q}_{\mathrm{Grav}}^{lj\,\prime
\prime \prime }+\frac{4}{3c^{4}}D_{e\&m}^{k\ \prime \prime }\overline{D}%
_{e\&m}^{j\ \prime \prime }  \label{P'} \\
&&+\frac{2}{135c^{6}}Q_{e\&m}^{jl\prime \prime \prime }\overline{Q}%
_{e\&m}^{lk\prime \prime \prime })-\frac{2\sqrt{2}}{15c^{5}}%
(Q_{e\&m}^{ik\prime \prime \prime }\overline{D}_{e\&m}^{k\prime \prime
}+D_{e\&m}^{k\prime \prime }\overline{Q}_{e\&m}^{ik\prime \prime \prime }) 
\nonumber
\end{eqnarray}

All the terms on the right-side are real\ and involve the products of
electric and magnetic type moments. Presumably (though we did not verify it)
the second and third terms arise from the Poynting vector hidden inside the
asymptotic Einstein-Maxwell equations. \ We refer to the right-side as the
(field) momentum recoil terms. \ By replacing the $P,\ $from Eq.(\ref{P}), ($%
P^{i}=M_{B}\xi _{R}^{i\prime }-\frac{2}{3}c^{-2}q^{2}\xi _{R}^{i\prime
\prime }\ $)$,\ $into Eq.(\ref{P'}), we obtain Newton's $2^{nd}$\ law:%
\begin{eqnarray}
M_{B}\xi _{R}^{i\prime \prime }\  &=&F^{i},  \label{2nd law} \\
F^{i} &=&\frac{2}{3c^{2}}q^{2}\xi _{R}^{i\prime \prime \prime
}-M_{B}^{\prime }\xi _{R}^{i\prime }\ -\frac{2\sqrt{2}}{15c^{5}}%
(Q_{e\&m}^{ik\prime \prime \prime }\overline{D}_{e\&m}^{k\prime \prime
}+D_{e\&m}^{k\prime \prime }\overline{Q}_{e\&m}^{ik\prime \prime \prime })
\label{F} \\
&&-i\epsilon _{ijk}(\frac{2}{15c^{4}}\boldsymbol{G}Q_{\mathrm{Grav}%
}^{kl\,\prime \prime \prime }\overline{Q}_{\mathrm{Grav}}^{lj\,\prime \prime
\prime }+\frac{4}{3c^{4}}D_{e\&m}^{k\ \prime \prime }\overline{D}_{e\&m}^{j\
\prime \prime }+\frac{2}{135c^{6}}Q_{e\&m}^{jl\prime \prime \prime }%
\overline{Q}_{e\&m}^{lk\prime \prime \prime }).  \nonumber
\end{eqnarray}%
The force consists of three parts: the classical radiation reaction term, a
rocket-like momentum loss term and the momentum recoil terms from the
fields. \ Note that one can have non-vanishing recoil arising from the
interaction of the electric quadrupole with an electric dipole, i.e., one
need not have magnetic and spin terms in order to have a field recoil force.

III. $l=2,$

For completeness we give, in outline form, the $l=2\ $harmonic equation, a
rather nasty non-linear relation for $\xi ^{ij},\ $coupled to $\xi ^{i}$.
Its explicit form does not shed much light on the physics, (the details just
need Clebsch-Gordon expansions on the right side): 
\begin{equation}
\xi _{R}^{ij\prime }-\frac{GM_{B}}{2c^{3}}\xi _{R}^{ij\prime \prime }+\frac{%
Gq^{2}}{9c^{7}}\xi _{R}^{ij\prime \prime \prime }=-\frac{1}{24}(2\eth %
L^{\prime }\ \overline{\eth }\overline{L}^{\prime }+Gc^{-4}\phi _{2}^{\ast 0}%
\overline{\phi }_{2}^{\ast 0})|_{l=2},\ \ \   \label{l=2}
\end{equation}%
with%
\begin{eqnarray}
\ \ L(u,\zeta ,\overline{\zeta }) &=&\xi ^{i}Y_{1i}^{1}-6\xi
^{ij}Y_{2ij}^{1}, \\
\ \ \phi _{2}^{\ast 0} &=&-2D_{e\&m}^{i\prime \prime }Y_{1i}^{-1}-\frac{%
\sqrt{2}}{12}Q_{e\&m}^{ij\prime \prime \prime }Y_{2ij}^{-1}.
\end{eqnarray}

\section{Summary and Discussion}

\subsection{Summary}

We have considered perturbations of the Reissner-Nordstrom metric that
remain in the class of algebraically special type II metrics. After all
radial integrations had been performed (already described in the literature)
there remained four relevant field equations (with a reality condition) for
four dependent variables, two Maxwell fields, ($\phi _{1}^{\ast 0},\phi
_{2}^{\ast 0}$), a Weyl tensor component, $\psi _{2}^{0}$ (or the mass
aspect, $\Psi $) and an angle field, $L$. The angle field describes the
direction that the shear-free null geodesic congruence associated with the
degenerate type II principal null vector field intersects future null
infinity, $\mathfrak{I}^{+}.\ $The independent variables are the Bondi
coordinates of $\mathfrak{I}^{+},\ $the time variable,$\ u,\ \ $and the
stereographic angles ($\zeta ,\overline{\zeta }$).\ 

The four field variables were expanded in spherical harmonics%
\begin{eqnarray}
L(u,\zeta ,\overline{\zeta }) &=&\eth _{(\tau )}G(\tau ,\zeta ,\overline{%
\zeta })=\xi ^{i}(\tau )Y_{1i}^{1}-6\xi ^{ij}(\tau )Y_{2ij}^{1}, \\
\phi _{1}^{\ast 0} &=&q+\phi _{1}^{\ast 0i}(\tau )Y_{1i}^{0}+\phi _{1}^{\ast
0ij}(\tau )Y_{2ij}^{0}, \\
\phi _{2}^{\ast 0} &=&-2c^{-2}D_{e\&m}^{i\prime \prime }(\tau )Y_{1i}^{-1}-%
\frac{\sqrt{2}}{12}c^{-3}Q_{e\&m}^{ij\prime \prime \prime }(\tau
)Y_{2ij}^{-1}, \\
\Psi &=&\Psi ^{0}(\tau )+\Psi ^{i}(\tau )Y_{1i}^{0}+\Psi ^{ij}(\tau
)Y_{2ij}^{0},
\end{eqnarray}%
(with the $u$\ behavior given parametrically by $u=G(\tau ,\zeta ,\overline{%
\zeta })=\tau -\frac{1}{2}\xi ^{i}(\tau )Y_{1i}^{0}(\zeta ,\overline{\zeta }%
)+\xi ^{ij}(\tau )Y_{2ij}^{0})$ and inserted into the first three of the
field equations. The first two determine the electric and magnetic dipole
and quadrupole moments in terms of $L\ $as

\begin{eqnarray}
D_{e\&m}^{i} &=&\text{\ }D_{e}^{i}+iD_{m}^{i}=q(\xi _{R}^{i}+i\xi
_{I}^{i})=q\xi ^{i}, \\
\text{\ }\ Q_{e\&m}^{ij\prime } &=&Q_{e}^{ij\prime }+iQ_{m}^{ij\prime }=-24%
\sqrt{2}cq\xi ^{ij}.
\end{eqnarray}%
while the third yields, from the real part, the Bondi mass and, from the
imaginary part, the conservation of angular momentum%
\begin{eqnarray}
P^{i} &=&M_{B}\xi _{R}^{i\prime }-\frac{2}{3}c^{-2}q^{2}\xi _{R}^{i\prime
\prime }\ , \\
J^{i\prime } &=&0, \\
J^{i} &=&M_{B}c\xi _{I}^{i}+\frac{2}{3}c^{-1}q^{2}\xi _{I}^{i\prime }.
\end{eqnarray}

The last of the field equations, (basically the evolution equation for $L),\ 
$contains the conservation of energy and momentum laws. \ 
\begin{eqnarray}
E^{\prime } &\equiv &M_{B}^{\prime }c^{2}=-\frac{1}{5}c^{-7}\boldsymbol{G}Q_{%
\mathrm{Grav}}^{ij\,\prime \prime \prime }\overline{Q}_{\mathrm{Grav}%
}^{ij\,\prime \prime \prime }-\frac{2}{3}c^{-3}D_{e\&m}^{i\prime \prime }%
\overline{D}_{e\&m}^{i\prime \prime } \\
&&-\frac{1}{180}c^{-5}Q_{e\&m}^{ij\prime \prime \prime }\overline{Q}%
_{e\&m}^{ij\prime \prime \prime },  \nonumber
\end{eqnarray}

\begin{eqnarray}
P^{i\prime } &=&-\frac{2\sqrt{2}}{15c^{5}}(Q_{e\&m}^{ik\prime \prime \prime }%
\overline{D}_{e\&m}^{k\prime \prime }+D_{e\&m}^{k\prime \prime }\overline{Q}%
_{e\&m}^{ik\prime \prime \prime }) \\
&&-i\epsilon _{ijk}(\frac{2\boldsymbol{G}}{15c^{4}}Q_{\mathrm{Grav}%
}^{kl\,\prime \prime \prime }\overline{Q}_{\mathrm{Grav}}^{lj\,\prime \prime
\prime }+\frac{4}{3c^{4}}\epsilon _{ijk}D_{e\&m}^{k\ \prime \prime }%
\overline{D}_{e\&m}^{j\ \prime \prime }+\frac{2}{135c^{6}}Q_{e\&m}^{jl\prime
\prime \prime }\overline{Q}_{e\&m}^{lk\prime \prime \prime }).  \nonumber
\end{eqnarray}

The momentum law easily is converted into Newton's 2$^{nd}\ $law:%
\[
M_{B}\xi _{R}^{i\prime \prime }\ =F^{i},
\]%
with

\begin{eqnarray}
F^{i} &=&\frac{2}{3c^{2}}q^{2}\xi _{R}^{i\prime \prime \prime
}-M_{B}^{\prime }\xi _{R}^{i\prime }\ -\frac{2\sqrt{2}}{15c^{5}}%
(Q_{e\&m}^{ik\prime \prime \prime }\overline{D}_{e\&m}^{k\prime \prime
}+D_{e\&m}^{k\prime \prime }\overline{Q}_{e\&m}^{ik\prime \prime \prime }) \\
&&-i\epsilon _{ijk}(\frac{2\boldsymbol{G}}{15c^{4}}Q_{\mathrm{Grav}%
}^{kl\,\prime \prime \prime }\overline{Q}_{\mathrm{Grav}}^{lj\,\prime \prime
\prime }+\frac{4}{3c^{4}}D_{e\&m}^{k\ \prime \prime }\overline{D}_{e\&m}^{j\
\prime \prime }+\frac{2}{135c^{6}}Q_{e\&m}^{jl\prime \prime \prime }%
\overline{Q}_{e\&m}^{lk\prime \prime \prime }),  \nonumber
\end{eqnarray}%
the force consisting of the radiation reaction term and different types of
recoil terms.

\subsection{Discussion}

The question immediately arises: how seriously should one take this? One's
first reaction might be: "yes, this is important. One sees, in a relatively
simple situation (type II metrics), so many of the fundamental relations of
classical mechanics. They arise with no model building and no
renormalization problems, just GR coupled to the Maxwell field and
constrained to type II. The associated field equations are interpretable as
the dynamics of a particle, yielding \textit{both its motion and the
evolution of its internal multipole structure}. These particles (or
space-times) are just time-dependent generalizations (that remain in the
algebraically special class) of the fundamental metrics; Schwarzschild,
Reissner-Nordstrom, Kerr-Newman." These observations certainly suggest that
they should be taken seriously.

On the other hand, for a variety of reasons, one must remain skeptical of
their physical importance. \ It is hard to imagine what type of physical
object could be identified with these type II particles. \ One also has the
immediate classical difficulty with the run-away behavior associated with
the radiation reaction force. (Could the momentum recoil force damp out the
run-away behavior? Unlikely?) One has the difficult conceptual issue of what
is the physical space in which these particles are moving, i.e., how do we
identify physically the complex position vector $\xi ^{a}(\tau ),$\ or,\
what is the physical or observational meaning to $\mathfrak{H}$-space and
the $\mathfrak{H}$-space world-line? $\ $On this last issue, there are
attempts to give (observational) physical meaning to $\xi ^{a}(\tau )\ $\cite%
{TT}, which are partially but not totally satisfactory. \ More specifically,
there is point of view that is dual to that of the complex world-line in $%
\mathfrak{H}$-space. By returning to the space-time interpretation of the
equations and considering the \textit{real shear-free, \underline{\textit{%
but twisting}}, null geodesic congruence} defined by the Goldberg-Sachs
theorem via type II metrics, we can interpret the "type II particle" as the
caustics of the congruence, i.e., the space-time points where $\rho
\rightarrow \infty ,$ 
\begin{eqnarray}
\rho  &=&-\frac{1}{r+i\Sigma }, \\
i\Sigma (u,\zeta ,\overline{\zeta }) &=&\frac{1}{2}\{\eth \overline{L}+L%
\overline{L}^{{\large \cdot }}-\overline{\eth }L-\overline{L}L^{{\large %
\cdot }}\},
\end{eqnarray}%
or where both $r\ \ $and $\Sigma \rightarrow 0.$\ Perturbatively, these
caustics have the form of a world-tube, $(S^{1}$x$R$),$\ $and can be
considered as the source of the type II metrics since the Weyl curvature and
the Maxwell field are singular on the tube. \ All the information contained
in the complex picture of $\mathfrak{H}$-space and its world-line can be
reconstructed via the real twisting congruence. Unfortunately, the geometry
of the complex point of view is simpler and far more attractive than the
real picture.

In spite of the difficulties, we find issues raised by these type II
particles to be rather fascinating. \ If they have no connection to physics,
why do they appear so connected; if they are connected, what is the precise
connection? It is hard to believe that all this is just a strange accident
or coincidence.

As a final comment, we note that the perturbations, with considerably more
effort, can and have been removed from their restriction to type II metrics%
\cite{LR}. Almost all asymptotically flat Einstein-Maxwell space-times have
been considered.\ Though there are similarities in the final results:
equations of motion, energy-momentum and angular momentum conservation laws,
etc., some of the difficulties disappear, others remain. \ It should be
emphasized that the results described here are from a very special example -
vastly simpler and straightforward - of the more general asymptotic
considerations.

\section{Acknowledgments}

We thank our many friends and colleagues who over the years have helped us
understand the algebraically special space-times. Special thanks go to the
Warsaw group of Andrzej Trautman, Pawel Nurowski, Jerzy Lewandowski, Jacek
Tafel and their students and the Oxford group of Roger Penrose, Lionel
Mason, Paul Tod with their students. Particular thanks must go to my former
student Timothy Adamo, now of Oxford University.

\end{document}